\documentclass{article}

\usepackage{amssymb}
\usepackage{amsthm}
\usepackage{amsmath}
\usepackage{epsfig}
\usepackage{mathptmx}      

\newcommand{\BD}{Becker-D{\"o}ring }
\renewcommand{\rho}{\varrho}
\renewcommand{\phi}{\varphi}
\newcommand{\nn}{\nonumber}
\newcommand{\pat}{\partial_t}
\newcommand{\ve}{\varepsilon}
\newcommand{\OT}{\Omega_T}
\newcommand{\io}{\int\limits_\Omega}
\newcommand{\suma}{\sum_{\alpha=1}^\nu}
\newcommand{\sumb}{\sum_{\beta=1}^\nu}
\newcommand{\dt}{\frac{\mathrm d}{\mathrm dt}}
\newcommand{\dx}{\,\mathrm{d}x}
\newcommand{\ovc}{\overline{\chi}}
\newcommand{\ovK}{\overline{K}}
\newcommand{\ovN}{\overline{N}}
\newcommand{\ovr}{\overline{\rho}}
\newcommand{\ovx}{\overline{x}}
\newcommand{\ovZ}{\overline{Z}}

\newcommand{\R}{{\mathbb R}}
\newcommand{\N}{{\mathbb N}}

\newtheorem{theorem}{Theorem}[section]
\newtheorem{lemma}[theorem]{Lemma}
\newtheorem{proposition}{Proposition}[section]
\theoremstyle{definition}

\numberwithin{equation}{section}

\begin{document}

\title{On a modified Becker-D{\"o}ring model for two-phase materials}

\author{Thomas Blesgen
\footnote{Department of Mathematics, Bingen University of Applied Sciences,
D-55411 Bingen, Germany, email: {\tt t.blesgen@th-bingen.de}}, 
Ada Amendola
\footnote{Department of Civil Engineering, University of Salerno,
Fisciano, Italy,
email: {\tt adaamendola1@unisa.it}},
Fernando Fraternali
\footnote{Department of Civil Engineering, University of
Salerno, Fisciano, Italy,
email: {\tt f.fraternali@unisa.it}}
}

\date{\today}

\maketitle

\begin{abstract}
This work reconsiders the \BD model for nucleation under isothermal conditions
in the presence of phase transitions.
Based on thermodynamic principles a modified model is derived
where the condensation and evaporation rates may depend on the phase parameter.
The existence and uniqueness of weak solutions to the proposed model are shown
and the corresponding equilibrium states are characterized in terms of response functions and constitutive variables.
\end{abstract}
\maketitle

\vspace*{4mm}
\hspace*{64pt}Reaction diffusion equations.

\vspace*{2mm}
{\it Submitted to Continuum Mechanics and Thermodynamics}

\section{Introduction}
\label{secintro}
Nucleation phenomena occur in various applications of great technological
importance. Here we only mention the formation of liquid droplets in
semiconductors as GaAs, \cite{Dreyer1,Dreyer2}, nucleation as final step of
recrystallisation, \cite{RRG2000}, and the importance of nucleation in steels,
e.g. due to cold rolling, accompanied by investigations of local stresses
or the chemical composition of the material. 
Nucleation and growth phenomena play a key role on the morphology and
macroscale properties of a variety of metallic structures, which include bulky
materials, thin films and nanoparticles (refer, e.g., to \cite{Haumesser20161}
and references therein).
The thermomechanical formation process of metallic polycrystals is strongly
affected by the orientation of the crystal lattice in the individual grains,
lattice curvature, dislocations, and the migration of grain boundaries
\cite{Ask2018}.
The mathematical modeling of such phenomena needs to account for the formation,
nucleation and growth of voids within the material through suitable
constitutive parameters \cite{Tutyshkin2017251}.

Becker and D{\"o}ring introduced their model in 1935 to predict the
formation of liquid droplets or bubbles in vapour in a stationary setup,
\cite{BD}. At the heart of this model, droplets may grow or shrink solely by
one mechanism, the attachment or detachment of one single molecule (monomer)
or atom. The problem was reformulated by Frenkel \cite{Frenkel}. Later, Burton
\cite{Burton} was the first to study the dynamical aspects of the system.
Recent numerical studies on this subject can be found in \cite{Hong2016655}
and references therein.

The metastability of the equations was studied in \cite{BCP}, \cite{Penrose89}.
The transition of the \BD model to the Lifshitz, Slyozov, and Wagner (LSW)
model is studied in \cite{Niethammer2,Niethammer3}, \cite{Penrose97}.
In \cite{BCP}, a Lyapunov functional for nucleation is introduced which is
not in accordance to the second law of thermodynamics. This led to a corrected
version of the standard model proposed by Dreyer and Duderstadt in
\cite{Dreyer2}. In \cite{HNN}, the existence and uniqueness of this
non-standard model were studied, where many results of \cite{BCP} could
be reused. A thorough survey on nucleation from the physical point of
view is \cite{Kash}, where a discussion of the driving forces, heterogeneous
nucleation models and an analysis of equilibrium states can be found, among
others.
The results of this article also have significant implications on the theory
and the understanding of dynamic recrystallization (DRX). Typically,
during DRX, new nuclei essentially free of dislocations are formed within
a highly demaged material. Evidently, the lattice orders of nuclei and
surounding substrate are very different. DRX is a of great technological
importance, especially during the hot and cold rolling of industrial steels, and
is still the subject of intense research. Here we only mention the
recent articles \cite{TVBK19,ZLCK19,Ble4,SSDW18} and references therein.

In this article we investigate the implications of the second law of
thermodynamics on the reaction scheme imposed by the \BD system.
In order for the equations to fulfill the second law of thermodynamics,
the reaction rates representing the condensation (attachment) and
evaporation (detachment) of monomers have to depend
on the order parameter $\chi$. The formalism goes back to \cite{Ble1}, where
the oxidation of a solid precipitate is formally modelled by chemical reactions.

The outline of this article is as follows.
In Section~\ref{secder} we introduce the new model and show the validity of the
second law of thermodynamics. In particular, this allows us to study the
dependence of the reaction rates on the order parameter in the generalized
\BD scheme.
In Section~\ref{secspecial} we discuss special cases of the new model and
compare with the non-standard \BD model introduced by Dreyer and Duderstadt.
Section~\ref{secexist} is dedicated to the existence and uniqueness of
weak solutions to the new model. Section~\ref{secstat} contains a short
characterization of the equilibrium states.
We end with some concluding remarks and an outlook.

\newpage
\section{Derivation of the model}
\label{secder}
We want to study nucleation in a two-phase material that is contained in 
$\Omega\subset\R^d$. The main application we have in mind is a solid such
as a single crystal or a polycrystal with nucleating droplets. Throughout
the text we shall assume that all nuclei have a perfect spherical shape.

For a given stop time $0<T<\infty$, let $\OT:=\Omega\times(0,T)$.
The shape of the two phases is determined by an order parameter
$\chi: \OT\to[0,1]$ which is an indicator function of one selected phase.
This ansatz gives rise to a diffuse interface model with mushy regions.
To simplify the thermodynamic reasoning, we assume that the temperature
$\theta$ is kept constant in $\OT$.

Let $Z_\alpha(x,t)\ge0$ denote the number of nuclei of size $\alpha$,
$\alpha\in\{1,2,\ldots,\nu\}$ at $x\in\Omega$ and time $0\le t\le T$, where
$\nu\in\N\cup\{+\infty\}$ specifies the largest occurring nucleus.
We use the notations
\[ Z\equiv Z(x,t)=(Z_\alpha(x,t))_{1\le\alpha\le\nu}
\equiv(Z_\alpha)_{1\le\alpha\le\nu} \]
and drop the argument $(x,t)$ when this is clear from the context.

In contrast to the original \BD model, $Z$ may depend on the spatial
position $x$ which means that the function
\[ \rho(Z(x,t)):=\suma\alpha Z_\alpha(x,t) \]
may vary over $\Omega$. In the above context, $Z_\alpha$ and $\rho(Z)$ define
the number density and the mass density of nucleating particles, respectively,
with regard to a characteristic volume. The intuitive physical picture is
that nucleation starts on a small microscopic length scale, whereas the phase
profile determined by $\chi$ is a macroscopic quantity.

The total number of nuclei is given by
\begin{equation}
\label{Ndef}
N(Z(x,t))=\suma Z_\alpha(x,t),\quad (x,t)\in\OT.
\end{equation}

The time evolution of $Z$ in $\OT$ is determined by the system of ordinary
differential equations
\begin{equation}
\label{BD}
\dt Z_\alpha(x,t)=J_{\alpha-1}(Z(x,t),\chi(x,t))-J_\alpha(Z(x,t),\chi(x,t)),
\quad1\le\alpha\le\nu
\end{equation}
with the initial condition
\[ Z(\cdot,0)=\tilde{Z} \quad\mbox{in }\Omega \]
and the fluxes
\begin{eqnarray}
\label{J0def}
J_0(Z(x,t),\chi(x,t)) &=& -\suma J_\alpha(Z(x,t),\chi(x,t)),\\
J_\alpha(Z(x,t),\chi(x,t)) &:=& \Gamma_\alpha^C(\chi(x,t))Z_\alpha(x,t)
-\Gamma_{\alpha+1}^E(\chi(x,t))Z_{\alpha+1}(x,t)\nn\\
&:=& K(x)R_\alpha(\chi(x,t))^{1/b_\chi}Z_\alpha(x,t) \nn\\
\label{rdef} && -R_{\alpha+1}(\chi(x,t))^{1/b_\chi}Z_{\alpha+1}(x,t),\qquad
1\le\alpha\le\nu.
\end{eqnarray}
If $\nu$ is finite, a further closedness condition is required. Here we only
consider
\begin{equation}
\label{BDlast}
\Gamma_\nu^C=\Gamma_{\nu+1}^E\equiv0.
\end{equation}
Other choices are discussed in \cite{Burton}, \cite{Penrose89}.

The equations (\ref{BD})-(\ref{BDlast}) (completed with a governing equation
for $\chi$ below) are related to the classical \BD system, \cite{BD}, but with
rates that may additionally depend on $x$
and on the phase parameter $\chi$. In the definition of $J_\alpha$, the
functions $\Gamma^C_{\alpha}(\chi)>0$, $\Gamma^E_{\alpha}(\chi)>0$
denote the {\it condensation} and the {\it evaporation} rates of a nucleus of
size $\alpha$. In (\ref{rdef}), $K\in L^\infty(\Omega)$ is a given positive
function, and
\begin{equation}
\label{bchidef}
b_\chi:=\chi b_1+(1-\chi)b_2
\end{equation}
for two positive constants $b_l$, $l=1,2$ that appear in the definition
(\ref{fl1}) of the free energy of phase $l$. The particular form of 
$\Gamma^C_{\alpha}$, $\Gamma^E_{\alpha}$ in (\ref{rdef}) will be worked out
later and can be justified a posteriori by thermodynamic considerations.
Eqn.~(\ref{J0def}) ensures that $\rho(Z)$ is conserved in $\OT$.

The free energy $F$ of the system is
\[ F=F(Z,\chi)=\io f(Z,\chi,\nabla\chi)\dx \]
with the free energy density $f=f(Z,\chi,\nabla\chi)$. For $f$ we make the
ansatz
\begin{equation}
\label{fdef}
f(Z,\chi,\nabla\chi)=\chi f_1(Z)+(1-\chi)f_2(Z)
+\theta\Big(W(\chi)+\frac{\gamma}{2}|\nabla\chi|^2\Big).
\end{equation}
Here, $f_l$ is the free energy density of phase $l$, $l=1,2$ and the last term
is due to the entropy of mixing. The scalar $\gamma>0$ determines the square
root of the thickness of the boundary layer between the two phases (assumed
constant here), and
\[ W(\chi):=\chi\ln(\chi)+(1-\chi)\ln(1-\chi) \]
is a double well potential.

In order to formulate the gouverning equation for $\chi$,
we need to smoothen the spatial variation of $x\mapsto Z(x,t)$ in $\Omega$.
To this end, we fix $\ve>0$ and choose a function $\phi\in C^\infty(\R^d)$ with
$\phi\ge0$ and $\int_{\R^d}\phi(y)\,dy=1$. We regularize $Z$ by the convolution
\begin{equation}
\label{convolute}
Z_\ve(x,t):=(Z(\cdot,t)*\phi_\ve)(x)=\int\limits_{\R^d}\phi_\ve(x-y)Z(y,t)\,dy
\end{equation}
with the kernel $\phi_\ve(x):=\ve^{-d}\phi(x/\ve)$. For the validity of
(\ref{convolute}), $Z(\cdot,t)\in L^1(\Omega)$ is required. So we postulate
for the initial value of (\ref{BD})
\begin{eqnarray*}
\hspace*{64pt} && \tilde{Z}\in L^1(\Omega),\quad\rho(\tilde{Z})>0
\mbox{ in }\Omega,\quad\\
&& \rho(\tilde{Z})\mbox{ is bounded uniformly in }\Omega,
\hspace*{105pt} ({\rm A1})\\
\hspace*{64pt} && \tilde{Z}_\alpha\ge0\mbox{ in }\Omega\mbox{ for }\alpha\ge1,
\quad\tilde{Z}_1>0\mbox{ in }\Omega.
\end{eqnarray*}

For the time evolution of $\chi$, a variety of different laws may be used,
as long as they are compatible with thermodynamics.
Here we choose the Allen-Cahn type formula
\begin{equation}
\label{AC}
\tau\pat\chi=-\frac{\partial F}{\partial\chi}(Z_\ve,\chi),
\end{equation}
where $\tau=\tau(\theta)$ is a positive constant that adjusts the time scale of
the propagation in $\chi$. The presence of $Z_\ve$ in Eqn.~(\ref{AC}) states
that the number of nuclei has to be integrated (summed) over a small spatial
region of size $\ve$. Hence, $\ve>0$ introduces a length scale into the model.

\begin{figure}[h!ptb]
\unitlength1cm
\begin{picture}(11.5,3.4)
\put(4.7,2.4){$\chi=0$}
\put(7.1,2.4){$\chi=1$}
\put(4.0,-0.20){\psfig{figure=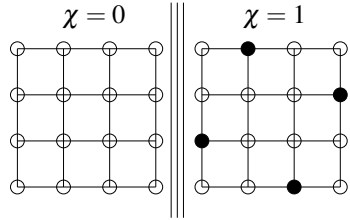,angle=0,width=4.5cm}}
\end{picture}
\caption{
\label{fig1}
Schematic illustration of a phase transition with two different lattice
geometries. Filled dots represent locked lattice sites.
Left a rectangular lattice with $b_1=1$, right a rectangular lattice
with $b_2=0.75$. The lines in the center represent the mushy region.}
\end{figure}

In the definition (\ref{fdef}), the functions $\{f_l\}_{l=1,2}$ are smooth
and convex and represent the specific free energy density of phase $l$.
A convenient choice for reactive systems is
\begin{equation}
\label{fl1}
f_l(Z):=k_B\theta\suma Z_\alpha\Big[b_l\ln\Big(
\frac{Z_\alpha}{N(Z)}\Big)+\frac{E_\alpha^l}{k_B\theta}\Big]\quad
\mbox{in }\OT,\quad l=1,2,
\end{equation}
where $k_B$ is the Boltzmann constant, $E_\alpha^l>0$ are enthalpic energy
terms, and $0<b_l\le1$ are constants representing the local lattice geometry of
phase $l$, $l=1,2$. In case of $b_l<1$, certain lattice sites are locked, e.g.
by impurities, geometrically necessary dislocations, or other
immobile constituents, see Figure~\ref{fig1}. In \cite{MDP06},
entropic terms of the form (\ref{fl1}) are derived from lattice models.

In the \BD model, the nucleation is modeled formally by reactions.
The reaction rates and the constants $E_\alpha^l$ are connected by the
{\it Arrhenius law}
\begin{equation}
\label{Arr}
R_\alpha(\chi)=\exp\Big(
\frac{\chi E^1_\alpha+(1-\chi)E^2_\alpha-E_A(\chi)}{k_B\theta}\Big),
\quad1\le\alpha\le\nu
\end{equation}
for an activation or sattle point energy $E_A(\chi)$ that has to be
exceeded to start the nucleation.

Exploiting the Arrhenius law (\ref{Arr}),
with $b_\chi$ as in (\ref{bchidef}), the free energy may be rewritten as
\[ F(Z,\chi)(t)= \io k_B\theta\!\sumb Z_\beta\Big[b_\chi
\ln\Big(\frac{Z_\beta R_\beta^{1/b_\chi}}{N(Z)}\Big)\!
+\!\frac{E_A(\chi)}{k_B\theta}\Big]+\theta\Big(W(\chi)
+\frac{\gamma}{2}|\nabla\chi|^2\Big)\dx. \]

A direct computation yields
\begin{eqnarray*}
\dt F(Z(t),\chi(t)) \!\!&=&\!\! \suma\frac{\partial F}{\partial Z_\alpha}
(Z,\chi)\pat Z_\alpha+\frac{\partial F}{\partial\chi}(Z,\chi)\pat\chi\\
\!\!&=&\!\! \io k_B\theta\!\!\suma\!\!\Big[b_\chi\ln\!\Big(
\frac{Z_\alpha R_\alpha^{1/b_\chi}}{N(Z)}\Big)\!+\!
\frac{E_A(\chi)}{k_B\theta}\Big]
\pat Z_\alpha\dx-\frac{1}{\tau}\Big(\frac{\partial F}{\partial\chi}(Z,\chi)
\Big)^2.
\end{eqnarray*}
Here we used that for any $1\le\alpha\le\nu$
\[ \sumb Z_\beta\frac{\partial}{\partial Z_\alpha}\Big[\ln\Big(
\frac{Z_\beta R_\beta^{1/b_\chi}}{N(Z)}\Big)\Big]=0. \]

If we resolve $\pat Z_\alpha(x,t)$ by the evolution law (\ref{BD}),
this becomes
\begin{eqnarray*}
\dt F(Z(t),\chi(t)) &=& \io\! k_B\theta\Big\{\!\Big(\!
-\!\suma J_\alpha(Z,\chi)
-J_1(Z,\chi)\Big)\!\Big[b_\chi\!\ln\!\!\Big(\frac{Z_1R_1^{1/b_\chi}}{N(Z)}\Big)
\!+\!\frac{E_A(\chi)}{k_B\theta}\Big]\\[-6pt]
&& \hspace*{8pt} +\!\sum_{\alpha=2}^\nu(J_{\alpha-1}(Z,\chi)\!-\!
J_\alpha(Z,\chi))\Big[b_\chi\!\ln\!\!\Big(\frac{Z_\alpha
R_\alpha^{1/b_\chi}}{N(Z)}\Big)
\!+\!\frac{E_A(\chi)}{k_B\theta}\Big]\!\Big\}\dx\\
&&\!\! -\frac{1}{\tau}\Big(\frac{\partial F}{\partial\chi}(Z,\chi)\Big)^2\\
&=& \io k_B\theta\suma J_\alpha(Z,\chi)\Big[b_\chi
\ln\!\Big(\frac{Z_{\alpha+1}R_{\alpha+1}^{1/b_\chi}}{Z_\alpha
R_\alpha^{1/b_\chi}}\frac{N(Z)}{Z_1R_1^{1/b_\chi}}\Big)
\!-\!\frac{E_A(\chi)}{k_B\theta}\Big]\dx\\
&& -\frac{1}{\tau}\Big(\frac{\partial F}{\partial\chi}(Z,\chi)\Big)^2.
\end{eqnarray*}
In glance of the thermodynamic structure of reactive systems,
this motivates to set
\begin{equation}
\label{r1set}
\frac{1}{K}\,\frac{Z_1R_1^{1/b_\chi}}{N(Z)}=\exp\Big(
\frac{-E_A(\chi)/b_\chi}{k_B\theta}\Big)\quad\mbox{in }\OT.
\end{equation}
For the existence proof of Section~\ref{secexist} we postulate the
following two conditions:

For each $1\le\alpha\le\nu$ and every $(x,t)\in\OT$ there exists
$0<\gamma_\alpha(x,t)<\infty$ such that
\begin{eqnarray*}
\hspace*{33pt} && \max\Big\{R_\alpha(\chi(x,t))^{1/b_1},\,
K(x)R_\alpha(\chi(x,t))^{1/b_1},\\
&& \hspace*{32pt} R_\alpha(\chi(x,t))^{1/b_2},\,
K(x)R_\alpha(\chi(x,t))^{1/b_2}\Big\}\le\gamma_\alpha(x,t)
\hspace*{43pt}({\rm A2})
\end{eqnarray*}
and for every $(x,t)\in\OT$ it holds
\[ \hspace*{92pt} \gamma_{\alpha+1}(x,t)\le\gamma_\alpha(x,t),\quad
\lim_{\alpha\to\infty}\frac{\gamma_\alpha(x,t)}{\alpha}=0.
\hspace*{65pt}({\rm A3}) \]

With (\ref{r1set}) and (\ref{rdef}), the final form of the free energy
inequality is
\begin{eqnarray}
\dt F(Z(t),\chi(t)) \!\!\!\!&=&\!\!\!\! \io k_B\theta\suma\Big(
KR_\alpha(\chi)^{1/b_\chi}Z_\alpha\!-\!R_{\alpha+1}(\chi)^{1/b_\chi}
Z_{\alpha+1}\Big)\nn\\
\label{Fest}
&& \hspace*{4pt} \times\, b_\chi\ln\Big(
\frac{R_{\alpha+1}(\chi)^{1/b_\chi}Z_{\alpha+1}}
{KR_\alpha(\chi)^{1/b_\chi}Z_\alpha}\Big)\dx
-\frac{1}{\tau}\Big(\frac{\partial F}{\partial\chi}(Z,\chi)\Big)^2.
\end{eqnarray}
Equality~(\ref{Fest}) immediately implies the validity of the second
law of thermodynamics, as in general $(B-A)\ln(A/B)\le0$ for arbitrary $A>0$
and $B>0$. So we unconditionally infer $\frac{d}{dt}F(Z(t),\chi(t))\le0$
as desired. Eqn.~(\ref{Fest}) also justifies the ansatz (\ref{rdef}).

\section{The relationship to the non-standard \BD model}
\label{secspecial}
In the special case of $b_1=b_2=1$, the system (\ref{BD})-(\ref{BDlast})
corresponds to the non-standard \BD model by Dreyer and Duderstadt,
\cite{Dreyer2}, that in contrast to the original \BD system satisfies the
laws of thermodynamics.

The condition $b_1=b_2=1$ generically holds for liquids and gases, but is also
fulfilled in solids if all lattice sites are freely accessible to nucleation.
In this case, we may choose $\Gamma_\alpha^C$, $\Gamma_{\alpha+1}^E$ independent
of $\chi$ to obtain $J_\alpha=J_\alpha(Z)$ for
\[ J_\alpha(Z)=\Gamma_\alpha^C(\cdot)Z_\alpha
-\Gamma_{\alpha+1}^E(\cdot)Z_{\alpha+1}\quad\mbox{in }\OT. \]
The equations (\ref{BD})-(\ref{BDlast}) decouple from $\chi$
and form a system of ordinary differential equations for the family of
unknowns $(Z_\alpha(x,t))_{1\le\alpha\le\nu}$ parameterized
by $x\in\Omega$.

The traditional \BD model uses constant condensation and evaporation rates,
\begin{equation}
\label{rconst}
\Gamma_\alpha^C(x,t)\equiv\Gamma_\alpha^C>0,\qquad
\Gamma_{\alpha+1}^E(x,t)\equiv\Gamma_{\alpha+1}^E>0.
\end{equation}
This constitutive assumption makes the number of nuclei independent of $x$, thus
$Z(t)=(Z_\alpha(t))_{1\le\alpha\le\nu}$. The equations (\ref{BD})-(\ref{BDlast})
then coincide with the \BD system, with
\[ \Gamma_\alpha^C=K\Gamma_\alpha^E\quad\mbox{for }1\le\alpha\le\nu, \]
which is a consequence of (\ref{rdef}).

With these simplifications, the approach (\ref{fl1}) relates to the choice
\[ f(Z(t))=k_B\theta\suma Z_\alpha(t)\ln\Big(
\frac{Z_\alpha(t)}{q_\alpha(t)N(Z(t))}\Big) \]
derived by Dreyer and Duderstadt in \cite{Dreyer2}, where
$q_\alpha=\exp(-\frac{E_\alpha}{k_B\theta})$. This is analogous to the formula
for $R_\alpha$ in (\ref{Arr}) if the activation energy $E_A$ is neglected.
The condition (\ref{r1set}) determines
$\Gamma_\alpha^C/\Gamma_\alpha^E$, 
whereas the condition
\[ \frac{N(Z(t))q_\alpha q_1}{Z_1(t)q_{\alpha+1}}=
\frac{\Gamma^E_{\alpha+1}}{\Gamma_\alpha^C} \]
chosen in \cite{Dreyer2} to guarantee the thermodynamic correctness of the
modified \BD system determines the ratio
$\Gamma_\alpha^C/\Gamma_{\alpha+1}^E$.

The ansatz (\ref{rdef}) is general enough to cover any of the commonly used
heuristic formulas for the evaporation and condensation rates like
\[ \Gamma_\alpha^C(t)=\alpha^A Z_1(t), \quad\Gamma_\alpha^E(t)=
\alpha^A\Big(C+\frac{D}{\alpha^B}\Big) \]
with constants $0\le A<1$, $0<B<1$, $C>0$, $D>0$.

Finally we mention that a formalism similar to (\ref{fl1}) is used in
\cite{Ble1}, where the thermodynamics of reactions accompanied by phase
transitions were studied and estimates similar to (\ref{Fest}) are found.
General reaction schemes and their asymptotic limits have also been
studied in \cite{Ble2}, where the reactions formally
model the generation and annihilation of vacancies in a solid
due to plastic effects accompanied by moving reconstitutive transition layers.

\section{Existence and uniqueness of weak solutions}
\label{secexist}
We proof existence and uniqueness of weak solutions to (\ref{BD})-(\ref{rdef}),
(\ref{AC}) for the most general case $\nu=\infty$.
By $C^k(I;\,S)$ we denote the
space of $k$-times continuously differentiable functions from an interval
$I\subset\R$ to a set $S$ and $H^{m,2}(\Omega)$ denotes the Sobolev space of
$m$-times weakly differentiable functions in the Hilbert space $L^2(\Omega)$.

The first step is to decouple (\ref{BD})-(\ref{rdef}) and slice the solution
$Z$ in the $x$-variable. This allows to apply the methods and results of
\cite{BCP}, \cite{HNN} for the \BD system.
For fixed $\ovx\in\Omega$ and given $\ovc(\ovx,\cdot)\in C^0([0,T];\,[0,1])$
we introduce the solution vector
\[ z(t)\equiv(z_\alpha(t))_{\alpha\in\N}\equiv z_{\ovx}(t):=Z(\ovx,t). \]
For the sliced system we seek solutions $t\to z_{\ovx}(t)$ in $C^0([0,T);\,X)$
with
\[ X:=\big\{(z_\alpha)_{\alpha\in\N}\;\big|\;\|z\|_X<\infty\big\},\quad
\|z\|_X:=\sum_{\alpha=1}^\infty\alpha|z_\alpha|. \]
We introduce the symbols
\begin{eqnarray*}
j_\alpha(z(t)) &:=& J_\alpha(Z(\ovx,t),\ovc(\ovx,t)),\\
r_\alpha(t) &:=& R_\alpha(\ovc(\ovx,t)),\\
k &:=& K(\ovx),\\
b(t) &:=& \ovc(\ovx,t)b_1+(1-\ovc(\ovx,t))b_2\ge\min\{b_1,b_2\}=:b_0>0.
\end{eqnarray*}
The system (\ref{BD})-(\ref{rdef}) becomes
\begin{eqnarray}
\label{bd1}
\dt z_\alpha(t) &=& j_{\alpha-1}(z(t))-j_\alpha(z(t)),\quad \alpha\ge1,\\
\label{bd2} j_0(z(t)) &=& -\sum_{\alpha=1}^\infty j_\alpha(z(t)),\\
\label{bd3} j_\alpha(z(t)) &=& k\,r_\alpha(t)z_\alpha(t)-r_{\alpha+1}(t)
z_{\alpha+1}(t),\quad\alpha\ge1.
\end{eqnarray}
As initial conditions we impose
\[ z(t=0)=\tilde{z}(0)\equiv\tilde{z}_{\ovx}(0):=\tilde{Z}(\ovx)=Z(\ovx,0). \]
Assumption~(A1) implies $\rho_0=\rho(\tilde{z})>0$ and $\tilde{z}_\alpha\ge0$
for all $\alpha\in\N$ and $\tilde{z}_1>0$.

With the above notations, the free energy density $f$ of the sliced system
becomes
\[ f(z)(t):=k_B\theta\sum_{\alpha=1}^\infty z_\alpha(t)\Big[b(t)\ln\Big(
\frac{z_\alpha(t)r_\alpha^{1/b(t)}}{N(z(t))}\Big)
+\frac{e_A(t)}{k_B\theta}\Big]+s_M, \]
where $e_A(t):=E_A(\ovc(\ovx,t))$ and $s_M$ is an integrating entropic constant.

Similar to \cite{BCP}, \cite{HNN} we can show:
\begin{proposition}
\label{Proposition1}
Let the assumptions (A2), (A3) hold. Then there exists a function
$z(t)=z_{\ovx}(t)\in C^0([0,T);\,X)$ which is the unique weak solution
of
\begin{eqnarray}
\dt z_\alpha(t) &=& j_{\alpha-1}(z(t))-j_\alpha(z(t)),\quad\alpha\ge2,\nn\\
\dt N_\alpha(z(t)) &=& j_{\alpha-1}(z(t)),\quad\alpha\ge2,\nn\\
\label{weaksol}
z_1(t_2)-z_1(t_1) &=& \int\limits_{t_1}^{t_2}j_0(z(t))\,dt,
\quad 0\le t_1,t_2<T.
\end{eqnarray}
Additionally it holds $z_\alpha(t)\ge0$ for all $\alpha\ge1$, $t\in[0,T)$, and
\begin{equation}
\label{cmass} \rho(z(t))=\rho(z(0))=\rho_0,\quad 0\le t<T,
\end{equation}
and for $0\le t_1\le t_2<T$ and a positive constant $C=C(\theta)$
\[ f(z(t_1))-f(z(t_2))\ge\frac{C}{\rho_0}\int\limits_{t_1}^{t_2}
\sum_{\alpha=1}^\infty|j_\alpha(z(t))|^2\,dt\ge0. \]
\end{proposition}

\noindent{\bf Remark.} Due to (\ref{weaksol}), $z$ is the {\it weak }
and not the strong solution to (\ref{bd1})-(\ref{bd3}).
\vspace*{2mm}

The proof of Proposition~\ref{Proposition1} is divided into several steps.
The equations (\ref{bd1})-(\ref{bd3}) are approximated by a finite system
of dimension $n$ that results from the infinite system by
neglecting all nuclei of size greater than $n$. So we consider
\begin{eqnarray}
\label{nbd1}
\dt z_\alpha^{(n)}(t) &=& j_{\alpha-1}(z^{(n)}(t))
-j_\alpha(z^{(n)}(t)),\quad2\le\alpha\le n-1,\\
\label{nbd2}
\dt z_1^{(n)}(t) &=& -j_1(z^{(n)}(t))
-\sum_{\alpha=1}^{n-1}j_\alpha(z^{(n)}(t)),\\
\label{nbd3}
\dt z_n^{(n)}(t) &=& j_{n-1}(z^{(n)}(t))
\end{eqnarray}
completed with the initial conditions
\[ z^{(n)}_\alpha(0)=\tilde{z}_\alpha^{(n)}\quad \mbox{for }1\le\alpha\le n. \]

\newpage
\noindent
It holds $\tilde{z}^{(n)}\to\tilde{z}$ in $X$ for $n\to\infty$, where
$\tilde{z}$ is the initial datum of the infinite system.

\vspace*{2mm}
On $X$ we define weakstar-continuous functionals
$N_\alpha(z):=\sum_{\beta=\alpha}^\infty z_\beta$. It clearly holds
$N(z)=N_1(z)$ and $z_\alpha(t)=N_\alpha(z(t))-N_{\alpha+1}(z(t))$.
One can formally show
\begin{eqnarray*}
\rho(z(t)) &=& \sum_{\alpha=1}^\infty N_\alpha(z(t)),\\
\dt N_\alpha(z(t)) &=& j_{\alpha-1}(z(t)),\quad \alpha\ge1.
\end{eqnarray*}

\vspace*{2mm}
\begin{lemma}
\label{Lemma1}
Let (A2), (A3) be satisfied. Then the following statements hold.

\noindent(i) For all $n\in\N$ there exists a solution
$z^{(n)}\in C^\infty([0,T);\,X)$
to (\ref{nbd1})-(\ref{nbd3}).

\noindent(ii) With $N_\alpha^{(n)}:=N_\alpha(z^{(n)})$,
$j_\alpha^{(n)}:=j_\alpha(z^{(n)})$,
$\rho^{(n)}:=\rho(z^{(n)})$, $f^{(n)}:=f(z^{(n)})$, the following statements
are valid:

\noindent(1) $\rho^{(n)}(t)=\rho^{(n)}(0)$.\\
(2) $\dt N_\alpha^{(n)}=j_{\alpha-1}^{(n)}(t)$ for all
$0\le t<T$ and all $1\le\alpha\le n$.\\
(3) There exists a constant $C=C(\theta)>0$ independent of $n$ such that
\begin{equation}
\label{fdecay}
f^{(n)}(t_1)-f^{(n)}(t_2)\ge\frac{C}{\rho^{(n)}}
\int\limits_{t_1}^{t_2}\sum_{\alpha=1}^{n-1}|j_\alpha^{(n)}(t)|^2\,dt
\end{equation}
for $0\le t_1\le t_2<T$.
\end{lemma}

\vspace*{2mm}
\begin{proof}
(i) This follows from the Picard-Lindel{\"o}f theorem.

\noindent(ii) We show only (\ref{fdecay}). The proof of the other statements is
similar.

A calculation as in (\ref{Fest}) yields
\[ \dt f(z^{(n)})(t)=-k_B\theta\,b(t)\sum_{\alpha=1}^{n-1}(c_\alpha-d_\alpha)
\big(\ln(c_\alpha)-\ln(d_\alpha)\big) \]
with $c_\alpha:=r_{\alpha+1}^{1/b(t)}z_{\alpha+1}^{(n)}(t)$,
$d_\alpha:=k\,r_\alpha^{1/b(t)}z_\alpha^{(n)}(t)$.

Let $\gamma_\alpha=\gamma_\alpha(\ovx,t)$.
It holds $z_\alpha^{(n)}\le\rho^{(n)}/\alpha$ and with the assumptions (A2),
(A3) we find 
$\max\{c_\alpha,d_\alpha\}\le\rho^{(n)}\gamma_\alpha/\alpha$ and therefore
\begin{eqnarray*}
k_B\theta\,b(t)(c_\alpha-d_\alpha)\big(\ln(c_\alpha)-\ln(d_\alpha)\big)
&\ge& k_B\theta\,b_0\frac{(c_\alpha-d_\alpha)^2}{\rho^{(n)}}
\frac{\alpha}{\gamma_\alpha}\\
&\ge& \frac{k_B\theta\,b_0C_1}{\rho^{(n)}}|j_\alpha^{(n)}|^2.
\end{eqnarray*}
In the last line we used (A3) which implies $\alpha/\gamma_\alpha\ge C_1>0$.
Setting $C:=k_B\theta\,b_0C_1$, the proof of (\ref{fdecay}) is complete.
\end{proof}

\newpage
\begin{lemma}
\label{Lemma2}
The following functions introduced in Lemma~\ref{Lemma1} are uniformly, i.e.
independently of $n$, bounded in $C^0([0,T))$.\\
(1) $z^{(n)}_\alpha$, $N_\alpha^{(n)}$, $j_\alpha^{(n)}$,
$\dot{z_\alpha}^{(n)}$, $\dot{N_\alpha}^{(n)}$ for $1\le\alpha\le n$.\\
(2) $\ddot{z_\alpha}^{(n)}$, $\ddot{N_\alpha}^{(n)}$, $\dot{j_\alpha}^{(n)}$
for $2\le\alpha\le n$.\\
(3) $j_0^{(n)}$, $\dot{j_1}^{(n)}$.
\end{lemma}

\vspace*{2mm}
\begin{proof} We demonstrate only (1).
We assume for simplicity $z_\alpha^{(n)}>0$ for all $\alpha$.
By direct computations we find
\begin{eqnarray}
\dt\sum_{\alpha=1}^n\alpha z_\alpha^{(n)}(t) &=& \sum_{\alpha=1}^n\alpha
\dot{z}_\alpha^{(n)}(t)\nn\\
&=& -\sum_{\alpha=1}^nj_\alpha^{(n)}(t)-j_1^{(n)}(t)+\sum_{\alpha=2}^n
\alpha\Big(j_{\alpha-1}^{(n)}(t)-j_\alpha^{(n)}(t)\Big)\nn\\
&=& -\sum_{\alpha=1}^nj_\alpha^{(n)}(t)-j_1^{(n)}(t)+\sum_{\alpha=2}^n
j_\alpha^{(n)}(t)+2j_1^{(n)}(t)\nn\\
\label{easy} &=& 0.
\end{eqnarray}
After integrating (\ref{easy}) w.r.t. $t$, we get
\[ z_\alpha^{(n)}(t)\le\sum_{\alpha=1}^n\alpha z_\alpha^{(n)}(t)
=\sum_{\alpha=1}^n\alpha z_\alpha^{(n)}(0)
=\sum_{\alpha=1}^n\alpha \tilde{z}_\alpha. \]
The term on the right is bounded independently of $n$.

The bounds on $\dot{z}_\alpha^{(n)}(t)$ can be derived directly,
\begin{eqnarray*}
|\dot{z}_\alpha^{(n)}(t)| \!\!&=&\!\! |j_{\alpha-1}^{(n)}(t)-j_\alpha^{(n)}(t)|
\le|j_{\alpha-1}^{(n)}(t)|+|j_\alpha^{(n)}(t)|,\quad 2\le\alpha\le n,\\
|\dot{z}_1^{(n)}(t)| \!\!&=&\!\! \Big|-j_1^{(n)}(t)-\sum_{\alpha=1}^n
j_\alpha^{(n)}(t)\Big|\le|j_1^{(n)}(t)|+\Big|\sum_{\alpha=1}^nj_\alpha^{(n)}(t)
\Big|.
\end{eqnarray*}
The bounds on $j_\alpha^{(n)}(t)$ can be derived as in Lemma~\ref{Lemma1}.
\end{proof}

\vspace*{2mm}
The uniform bounds of Lemma~\ref{Lemma2} as a consequence of the
Arzel{\`a}-Ascoli theorem permit to pass to the limit $n\to\infty$.
The solution $z$ in Proposition~\ref{Proposition1} is the limit of
$(z^{(n)})_{n\in\N}$.

The uniqueness of $z$ follows from a Gronwall estimate and the conservation of 
mass (\ref{cmass}) as outlined in \cite{HNN}. This completes the proof of
Proposition~\ref{Proposition1}. \qed

\begin{proposition}
\label{Proposition2}
Under the assumption (A1) there exists a unique solution $\chi$ to the
regularized Allen-Cahn equation (\ref{AC}) that satisfies

\noindent(i) $\chi\in C^{0,\frac{1}{4}}([0,T];\,L^2(\Omega))$,

\noindent(ii) $\pat\chi\in L^2(\OT)$,

\noindent(iii) $\ln(\chi)\in L^1(\OT)$ and $0<\chi<1$ almost everywhere.
\end{proposition}

\vspace*{2mm}
\begin{proof}
The statements (i)-(iii) follow from well-established
existence and uniqueness results of the Allen-Cahn equation, see for
instance \cite{Ble3}, provided we can show that the convolution
(\ref{convolute}) is well defined, i.e. if
\[ Z(\cdot,t)\in L^1(\Omega)\quad\mbox{for }0\le t<T. \]
For $t=0$, this follows from (A1). For $t>0$, it is sufficient to show
that there exists a function $g\in L^1(\Omega)$ such that
$Z(\cdot,t)$ is measurable and
\[ |Z_\alpha(x,t)|\le g(x)\quad\mbox{for almost every }x\mbox{ in }\Omega \]
and any $\alpha\ge1$.
But the last follows from (\ref{cmass}) and (A1), since
\[ Z_\alpha(x,t)\le\rho(z_x(t))=\rho(\tilde{z}_x)=\rho(\tilde{Z}(x))\le C. \]
As it is evident that $Z(\cdot,t)$ is measurable, the proof is complete.
\end{proof}

With the Propositions~\ref{Proposition1} and ~\ref{Proposition2}, the
subsequent theorem is now evident.
\begin{theorem}
\label{Theorem1}
Let the assumptions (A1)-(A3) be fulfilled. Then the system
(\ref{BD})-(\ref{AC}) possesses a unique
weak solution $(Z,\chi)$, where $Z(x,\cdot)\in C^0([0,T);\,X)$ for almost
every $x\in\Omega$ fulfills the properties stated in
Proposition~\ref{Proposition1} and where
$\chi\!\in\!C^0([0,T];L^2(\Omega))$
satisfies the properties stated in Proposition~\ref{Proposition2}.
\end{theorem}

\section{Characterization of the equilibrium states}
\label{secstat}
In this section we characterize the equilibrium states
$(\ovZ,\ovc)$. Stationarity in $\chi$ requires 
\[ \frac{\partial F}{\partial\chi}(\ovZ,\ovc)=0 \]
and due to the gradient term
$\frac{\gamma}{2}|\nabla\chi|^2$ in $F$ this yields $\ovc\equiv{\rm const}$ in
$\Omega$. This in turn shows $\ovZ_\alpha\equiv{\rm const}$ in $\Omega$ for
every $\alpha$. Clearly, $(\ovZ\equiv0,\ovc)$ is always an equilibrium state.
For $\ovZ$ not completely vanishing, we prescribe the local mass $\ovr>0$ and
want to prove that there exists an equilibrium state with $\ovr(\ovZ)=\ovr>0$.
Stationarity in $Z$ implies $J_\alpha(\ovZ,\ovc)=0$. With (\ref{rdef}) we find
\begin{equation}
\label{recurse}
\ovK\Gamma_\alpha^E\ovZ_\alpha=\Gamma_{\alpha+1}^E\ovZ_{\alpha+1},
\end{equation}
where $K(x)\equiv\ovK$ in $\Omega$ is assumed.

The rates $\Gamma_\alpha^E$, $\Gamma_\alpha^C$ are always positive numbers.
Here we make the stronger assumption that they are bounded away from zero
uniformly in $\alpha$, i.e. for a constant $c_0>0$
\[ \hspace*{91pt} \Gamma_\alpha^E(\ovc)=R_\alpha^E(\ovc)\ge
c_0\quad \mbox{for all }\alpha\in\N.\hspace*{86pt}\rm{(A4)} \]
Iterating (\ref{recurse}) yields
\begin{equation}
\label{Zbardef}
\ovZ_\alpha=\ovN s(\Gamma_\alpha^E)^{-1}\ovK^\alpha. 
\end{equation}
Here, $s=s(\ovc)=\exp(-E_A(\ovc)/(k_B\theta b_{\ovc}))$, since by
(\ref{r1set}) it holds $\ovZ_1\Gamma_1^E(\ovc)=\ovN Ks(\ovc)$.

From $\ovN(\ovZ)=\ovN$, $\rho(\ovZ)=\ovr$ we obtain
\[ \ovN s\sum_{\alpha=1}^\infty(\Gamma_\alpha^E(\ovc))^{-1}\ovK^\alpha=\ovN,
\quad \ovN s\sum_{\alpha=1}^\infty\alpha(\Gamma_\alpha^E)^{-1}\ovK^\alpha=\ovr.
\]
In order to derive a condition on $\ovK$, this is re-written in the form
\[ \tilde{f}(\ovK)=1,\quad \ovN=\frac{\ovr}{\tilde{g}(\ovK)} \]
with
\[ \tilde{f}(K):=s\sum_{\alpha=1}^\infty(\Gamma_\alpha^E)^{-1}K^\alpha, \quad
\tilde{g}(K):=s\sum_{\alpha=1}^\infty\alpha(\Gamma_\alpha^E)^{-1}K^\alpha. \]
Assumption~(A4) ensures the convergence of both series with radius of
convergence $1$. The function $\tilde{f}$ is continuous, strictly increasing in
$[0,1]$, and fulfills $\tilde{f}(K)\ge s(\Gamma_1^E)^{-1}K$. Therefore
$\ovK\in[0,\min\{1,\Gamma_1^E/s\}]$ if and only if $\tilde{f}(1)\ge 1$.
To fulfill the second condition $\rho(\ovZ)=\ovr>0$,
we require $\tilde{g}(\ovK)<\infty$. But from $\tilde{f}(1)>1$ we infer $\ovK<1$
and thus $\tilde{g}(\ovK)<\infty$. So we arrive at the following sufficient
condition of an equilibrium state:
\[ \hspace*{77pt} \tilde{f}(1)>1, \quad\mbox{or } (\tilde{f}(1)=1\mbox{ and }
\tilde{g}(1)<\infty). \hspace*{80pt}(EQ) \]

\vspace*{2mm}
\begin{proposition}
\label{Proposition3}
Let (EQ) and Assumption~(A4) hold. Then, for any $\ovr>0$, there exists an
equilibrium state $(\ovZ,\ovc)$ of (\ref{BD})--(\ref{rdef}), (\ref{AC}) with
$\rho(\ovZ)=\ovr$. Furthermore,

\noindent(a) $(\ovZ_\alpha)_{\alpha\ge1}$ and $\ovc$ are constant functions in
$\Omega$ which satisfy
\begin{eqnarray*}
\frac{\partial F}{\partial\chi}(\ovZ,\ovc) &=& 0,\\
J_\alpha(\ovZ,\ovc) &=& 0 \quad \mbox{for }\alpha\ge1.
\end{eqnarray*}
(b) The function $K$ fulfills $K\equiv\ovK$ in $\Omega$ with a unique
constant $\ovK\in(0,1]$ such that
\[ \tilde{f}(\ovK)=1. \]
(c) $\ovZ_\alpha$ is given by (\ref{Zbardef}) for all $\alpha\in\N$,
i.e. $\ovZ_\alpha=\ovN s (\Gamma_\alpha^E)^{-1}\ovK^\alpha$, where
$\ovN=N(\ovZ)=\ovr/\tilde{g}(\ovK)$.
\end{proposition}

\section{Concluding remarks}
\label{secdiss}
In this article we derived a generalized \BD model for nucleation in the
presence of phase transitions which respects the second law of thermodynamics,
as opposed to the original formulation \cite{BD}. A key feature of the
proposed model relies on the prediction that condensation and evaporation rates
depend on the phase parameter $\chi$.
The mathematical formulation of this model led us to prove existence and
uniqueness of weak solutions, and to characterize the equilibrium states
of the system.

A limitation of the developed theory, as for the \BD equations in general,
is that it is a system of {\it ordinary} differential equations. In addition,
the dependence on the spatial coordinate $x$ is not known. 
We plan to further investigate this issue through future work, which will also
be aimed at numerically studying the dependence on the parameters and their
spatial variation in more detail. Such a study will employ both deterministic
and statistical approaches, and will be focused on the analysis of
time-dependent properties and metastability, since available literature results
for \BD models are still incomplete in terms of convergence rates, coarsening
effects and evolution of large clusters as time goes to infinity
(refer, e.g., to \cite{Hingant2017175} and references therein).

\vspace*{6mm}
{\bf Acknowledgements}
This article was written while TB visited the Hausdorff Research
Institute for Mathematics (HIM), University of Bonn, in 2019. This visit was
supported by the HIM. TB gratefully acknowledges both this support and the
hospitality of HIM. AA and FF gratefully acknowledge financial support from the
Italian Ministry of Education, University and Research (MIUR) under the
`Departments of Excellence' grant L.232/2016.

\end{document}